\documentclass[aps,nofootinbib,a4wide,prd]{revtex4-2}

\usepackage{hyperref}
\usepackage{braket}
\usepackage{amssymb}
\usepackage{amsmath}
\usepackage{amsthm}
\usepackage{dsfont}
\usepackage{amscd}
\usepackage[all,arc,knot,matrix]{xy}

\usepackage{graphicx}
\usepackage{mathpazo}
\usepackage{CJK}

\theoremstyle{definition}

\theoremstyle{remark}

\usepackage{braket}
\usepackage{graphicx}
\usepackage{subfig}

\usepackage{hyperref}
\hypersetup{
  colorlinks=true,
  urlcolor=black,
  citecolor=blue
}

\begin{document}

\title{Semiclassical multi-mode squeezed states in harmonic cosmology}

\author{Bekir Bayta\c{s}}
\email{bekirbaytas@iyte.edu.tr}
\affiliation{Department of Mathematics,
\.{I}zmir Institute of Technology,
G\"ulbah\c{c}e, Urla 35430, \.{I}zmir, Turkey}

\date{\today}

\begin{abstract}
Harmonic cosmology is an algebraic framework based on the $\mathfrak{su}(1,1)$ Lie algebra that underlies the effective dynamics of loop quantum cosmology and group field theory. Focusing on the multimode bosonic realization of $\mathfrak{su}(1,1)$, we construct multimode squeezed states in harmonic cosmology and analyze their semiclassical properties by computing the relative uncertainties of the volume operator at the kinematical level. In contrast to single-mode squeezed vacua, which invariably exhibit large quantum fluctuations, we find that the relative uncertainty of the volume can be made arbitrarily small when the number of modes is large, even without displacement. Our results demonstrate that a kinematical squeezed vacuum state with a large number of independent modes can behave semiclassically providing insights into the quantum nature of bouncing cosmologies and the emergence of a classical universe.
\end{abstract}


\maketitle

\section{Introduction}

Semiclassical approximations are often useful when describing physical systems with subdominant quantum effects. There are two main implementations of such approximations: choosing a specified set of states and observables with semiclassical features (canonical formalism) or stationary phase solutions (path-integral formalism). The well-known application of the former implementation is realizing a classical electromagnetic field approximated by coherent or squeezed states in quantum optics~\cite{Walls83,Loudon2000,Zoller2004}.  The main importance of these states is that they provide a correspondence between expectation values and classical field configurations, thereby enabling a transition from the quantum to the classical domain.

In quantum gravity, quantum effects become highly dominant when the curvature or matter density reaches Planckian values; otherwise, a semiclassical spacetime description is adequate for phenomena in which quantum superpositions of spacetimes are hardly observable. In most quantum gravity theories, the underlying mathematical structures are highly non-trivial in the absence of symmetry considerations. Consequently, quantum cosmological models play a fundamental role in the study of quantum gravity by incorporating symmetry reductions that provide a more tractable framework than the full theory~\cite{BojLQC15}. The resulting minisuperspace models typically possess a finite number of degrees of freedom, and the challenge is to extract a classical, smoothly evolving spacetime from the solutions of the quantum constraints.

In the early studies of quantum cosmology, specifically Wheeler-DeWitt (WD) cosmology~\cite{DeWitt67}, semiclassical spacetimes were described by a wave function in a highly oscillating phase, not localized in the configuration space of geometries; this is the WKB approximation, as opposed to using coherent states sharply peaked around classical solutions. This approximation was heavily used in cosmological perturbation theory, where inflationary cosmology, formulated in the context of quantum field theory in curved spacetime, emerges from the semiclassical limit of quantum cosmology~\cite{Hawking85, Kiefer87}. Although highly successful, the WKB method relies on a rapidly varying phase and does not yield a localized wave packet in configuration space. The interpretation of a classical background in this approach is therefore less direct than in approaches that employ genuine minimum-uncertainty states. Coherent states, on the other hand, naturally yield sharply peaked Gaussian wave functions and permit a more transparent passage to the classical limit.

Recent advancements in quantum cosmology have been significantly informed by developments in loop quantum gravity (LQG)~\cite{AL04, Thi07, Rovelli2004}, which is one of the leading approaches to quantum gravity and offers a non-perturbative and background-independent setting for the unification of general relativity and quantum physics. There are well-studied proposals for coherent states in the full LQG~\cite{Thi01, Thi01-2, Freidel11, Livine20}, where they exhibit the expected feature of being sharply peaked around a given classical configuration with small uncertainties. Rather than considering the full LQG, interest in the cosmological symmetry reduction of LQG gave rise to the theory of loop quantum cosmology (LQC)~\cite{BojLQC08,AS11}, which predicts scenarios for the early universe, including the resolution of the big-bang singularity and the emergence of a cosmic bounce in the evolutionary dynamics. Generalized classes of semiclassical states in LQC have also been extensively studied~\cite{Corichi11, AS15, Boj07b}. In particular, the effective dynamics of LQC is derived by evaluating the Hamiltonian constraint on states sharply peaked on a classical trajectory~\cite{Taveras:2008ke}.

However, the predictions, e.g.\ the big bounce, are not exclusive to LQC and have also been realized within other alternative approaches to quantum cosmology~\cite{Ambjorn05, Bonanno02}. One of the approaches, whose canonical formulation (Fock space quantization) resembles that of LQG (spin-network states), is the group field theory (GFT) approach~\cite{Freidel05,Oriti17,Oriti16,Oriti07}. Semiclassical features used in GFT, mostly relevant to quantum cosmological models~\cite{CalGieOri12,GieOriSin13}, are described in terms of condensate states of quantum geometry~\cite{CalGieOri14, GieOri14, Ori17-1,Gie15,Gielen16,Calcinari:2026hbm}. These condensate states are characterized using Fock coherent states built on the annihilation and creation operators of the theory; many cosmological applications have employed this type of semiclassical coherent states~\cite{Oriti16-1,Oriti16-2,Gielen23}. In this setting, the volume operator is expressed as a combination of number operators, and its fluctuations encode the quantum properties of the geometry.

All these convergences in predictions prompt the question of what fundamental aspects of quantum gravity underlie these novel results across different frameworks. Reference~\cite{BBC19} revealed that the effective classical dynamics derived from LQC and GFT are, in fact, distinct manifestations of a unifying algebraic structure known as harmonic cosmology~\cite{Boj07,Boj08}. The central algebraic framework of harmonic cosmology is the $\mathfrak{su}(1,1)$ Lie algebra, which is isomorphic to $\mathfrak{sp}(2,\mathbb{R})$ and $\mathfrak{sl}(2,\mathbb{R})$. It has been demonstrated that the effective cosmological dynamics in LQC and GFT correspond to the canonical and bosonic realizations, respectively, of $\mathfrak{su}(1,1)$ harmonic cosmology. Thus, the $\mathfrak{su}(1,1)$ Lie algebra emerges as a fundamental structure underlying quantum cosmological models with bouncing behaviors. This assertion is further corroborated by the Complexifier-Volume-Hamiltonian (CVH) algebra in LQC~\cite{BL17}, which also exhibits an $\mathfrak{su}(1,1)$ Lie algebraic structure~\cite{Bodendorfer20}. The appearance of this algebra is not accidental: in homogeneous and isotropic models, the gravitational canonical pair and the matter degrees of freedom typically combine into a single pair of harmonic-oscillator-like variables, making the $\mathfrak{su}(1,1)$ structure manifest.

The $\mathfrak{su}(1,1)$ Lie algebra, along with its associated $\mathrm{SU}(1,1)$ Lie group, has proven instrumental in constructing algebraic coherent and squeezed states in quantum optics. In the discrete series representations, Perelomov generalized coherent states are obtained by acting with the group displacement operator on a lowest-weight state, and they inherently incorporate squeezing. In the bosonic realization, these correspond to the well-known single-mode and two-mode squeezed vacuum states, generated by the squeeze operator that is bilinear in creation and annihilation operators. Inspired by these applications, one can construct coherent and squeezed states within $\mathfrak{su}(1,1)$ quantum cosmology, providing insights into its semiclassical limit beyond the effective dynamics~\cite{LM12}. In the context of GFT, the construction of squeezed states, coherent states, and squeezed coherent thermal states has been extensively explored~\cite{AGW18,GP20,CG23}, utilizing the $\mathfrak{su}(1,1)$ Lie algebra. Notably, the relative uncertainties of the volume operator in these states have been computed to characterize their semiclassical properties. Unlike coherent states, single-mode squeezed states---such as dipole condensate states---exhibit large relative uncertainties in the volume operator and are therefore not semiclassical. By displacing and mixing these squeezed states, one obtains squeezed coherent thermal states that include semiclassical states characterized by small relative uncertainties~\cite{CG23}. Furthermore, the introduction of thermal states in GFT~\cite{AK20,Guo21} enables the mixing of single-mode squeezed states into generalized Gaussian states, including mixed states. These results highlight that, while a single uncorrelated squeezed mode alone lacks a well-defined classical geometry, appropriate statistical mixtures or displacements can restore semiclassicality.

In this paper, we focus on the generalization of single-mode squeezed states, which involves considering multimode squeezed states. In the case of $\mathrm{SU}(1,1)$ squeezed states, their multimode generalizations retain the $\mathfrak{su}(1,1)$ multimode bosonic realization~\cite{LL93,LS93,FF97}, suggesting that multimode GFT squeezed states may also constitute a bosonic realization of $\mathfrak{su}(1,1)$ harmonic cosmology. In such a realization, the algebra is represented as a sum over independent two-mode squeezing operators, leading to states that can be interpreted as correlated pairs of quanta across multiple modes~\cite{Calcinari:2026hbm}. Firstly, we undertake a concrete study of the multimode bosonic realization of $\mathfrak{su}(1,1)$ harmonic cosmology. Then, following the framework established in GFT cosmology~\cite{AGW18,GP20,CG23}, we explicitly compute the relative uncertainties of the volume operator for such multimode squeezed states. We restrict the semiclassical analysis of the multimode squeezed states to the kinematical level of the harmonic cosmology. Our main focus is to show that, with an increasing number of modes and a suitable distribution of squeezing parameters, the relative volume uncertainty can be significantly suppressed---contrary to the high quantum fluctuations of single-mode squeezed states---thus endowing multimode squeezed states with a robust semiclassical interpretation.

The paper is organized as follows. In Sec.~\ref{SII}, we review classical harmonic cosmology and its $\mathfrak{su}(1,1)$ algebraic structure, including the bosonic realizations and the connection to Friedmann-Lema\^itre-Robertson-Walker (FLRW) cosmology. Section~\ref{sec:quantum} presents the quantum theory, focusing on the multimode bosonic realization and the construction of multimode squeezed states. We then compute the expectation value and variance of the volume operator in these states. In particular, Sec.~\ref{sec:semiclassical} discusses the semiclassicality of the states and compares with single-mode results. Section~\ref{sec:conclusion} summarizes our findings and outlines future directions. Appendix~\ref{app:multimode} provides details on the multimode bosonic realization of $\mathfrak{su}(1,1)$.

\section{Harmonic cosmology}\label{SII}

In this section, we introduce the primary algebraic features of harmonic cosmology. First, we review the set of observables and the Hamiltonian of the model, together with the algebra that governs its dynamics. We then give an explicit description of the bosonic realizations of harmonic cosmology and finally discuss the connections to the FLRW model of classical cosmology and to LQC.

\subsection{Algebraic setting of harmonic cosmology}
A harmonic cosmological model~\cite{Boj07,Boj08} can be regarded as a classical dynamical system characterized by a set of observables $(V,J,\bar{J})$ that form a Poisson algebra isomorphic to the Lie algebras $\mathfrak{su}(1,1) \sim \mathfrak{sp}(2,\mathbb{R}) \sim \mathfrak{sl}(2,\mathbb{R})$:
\begin{align}
\{V,J\} = i \tilde{\delta} J\,, \quad \{V, \bar{J}\} =- i \tilde{\delta} \bar{J}\,, \quad \{J,\bar{J}\} = -2 i \tilde{\delta} V,
\label{eq:1Poisson}
\end{align}
with a complex variable $J$ and a fixed real parameter $\tilde{\delta}$. In terms of the real variables $V$, $J_+ \equiv \mathrm{Re}J$, $J_- \equiv \mathrm{Im}J$, the algebra~\eqref{eq:1Poisson} becomes
\begin{align}
\{V,J_{\pm}\} = \mp \tilde{\delta} J_{\mp}\,, \quad \{J_+, J_-\} = \tilde{\delta} V \,.
\label{eq:2Poisson}
\end{align}
The Casimir $\mathcal{C}$ of the Lie algebra \eqref{eq:1Poisson} (or \eqref{eq:2Poisson}) is given by the norm of the vector $(J_+,J_-,V)$ in three-dimensional Minkowski space $\mathbb{R}^{2,1}$:
\begin{equation}
\mathcal{C} \equiv V^2 - J^2_+ - J^2_- = V^2 - \frac{1}{2}(J \bar{J} + \bar{J} J),
\label{eq:Casimir}
\end{equation}
so that the vector $(J_+,J_-,V)$ is timelike for $\mathcal{C}>0$ and null for $\mathcal{C}=0$.

The Hamiltonian of the system, $H^{\tilde{\delta}}_{\varphi}$, which is an element of the algebra~\eqref{eq:2Poisson}, is defined by
\begin{equation}
H^{\tilde{\delta}}_{\varphi} := \frac{1}{\tilde{\delta}} \frac{1}{2 i}(J-\bar{J}) = \frac{1}{\tilde{\delta}} \, J_- \,.
\label{eq:hamiltonian}
\end{equation}

The subscript $\varphi$ indicates that the Hamiltonian generates translations in the parameter $\varphi$; it does not imply that $H^{\tilde{\delta}}_{\varphi}$ is a function of $\varphi$, since $H^{\tilde{\delta}}_{\varphi}$ contains only the phase-space observables $J$ and $\bar{J}$, which are independent of $\varphi$. Thus the Hamiltonian is autonomous\footnote{The Hamiltonian that drives evolution with respect to a relational parameter (here $\varphi$) is a constant of motion, ensuring that the equations of motion are translation-invariant.}. The equations of motion for $(J_+,V)$ are then obtained from the Poisson brackets with $H^{\tilde{\delta}}_{\varphi}$:
\begin{equation}
\frac{d V}{d \varphi} = \{V, H^{\tilde{\delta}}_{\varphi}\} = J_+ \,, \quad \frac{d J_+}{d \varphi} = \{J_+, H^{\tilde{\delta}}_{\varphi}\} = V\,.
\label{eq:eqm-harmonic}
\end{equation}
When $\mathcal{C}=0$, the solutions of these equations are
\begin{align}
V(\varphi) = \tilde{V} \cosh(\tilde{\delta}(\varphi-\varphi_0))\,, \quad J_+(\varphi) = \tilde{V} \sinh(\tilde{\delta}(\varphi-\varphi_0))\,,
\label{eq:solns}
\end{align}
where $\tilde{V}$ is the minimal value of $V(\varphi)$, reached at $\varphi = \varphi_0$. The flow generated by $H^{\tilde{\delta}}_{\varphi}$ clearly shows that the Hamiltonian acts as a pure boost in the $(J_+,V)$ plane.

When the Casimir $\mathcal{C}$ is non-negative -- i.e. the vector $(J_+,J_-,V)$ is timelike or null -- the solution~\eqref{eq:solns} implies
\begin{equation}
\mathcal{C} = \tilde{V}^2 - (\tilde{\delta} H^{\tilde{\delta}}_{\varphi})^2 \geq 0 \,.
\end{equation}
The condition $\mathcal{C}>0$ introduces a non-zero kinematical minimal volume $V_{\mathrm{min}}$ that separates from the dynamical contribution, as
\begin{align}
V|_{\varphi = \varphi_0} = \tilde{V} := \sqrt{(\tilde{\delta} H^{\tilde{\delta}}_{\varphi})^2 + V^2_{\mathrm{min}}} \, \geq \, V_{\mathrm{min}} > 0\,.
\end{align}
If instead $\mathcal{C}=0$, the kinematical contribution vanishes, giving $\tilde{V} = \pm \tilde{\delta} H^{\tilde{\delta}}_{\varphi}$.

Because the observables $(V,J,\bar{J})$ form a Lie algebra, it is natural to explore various realizations of the algebra~\cite{BBC19}. In the following we shall employ a specific multimode bosonic realization of $\mathfrak{su}(1,1)$~\cite{LL93,LS93,FF97}, defined in terms of a set of complex variables $(\bar{z}_i, z_i)$ with $i=1,\dots,N$:
\begin{align}\label{9}
V = \frac{\tilde{\delta}}{4} \, \sum_{i=1}^N \, (\bar{z}_i z_i + z_i \bar{z}_i) \,, \quad
J = \frac{\tilde{\delta}}{2} \, \sum_{i,j=1}^N \, \beta_{ij} \, \bar{z}_i \bar{z}_j \,, \quad
\bar{J} = \frac{\tilde{\delta}}{2} \, \sum_{i,j=1}^N \, \bar{\beta}_{ij} \, z_i z_j \,,
\end{align}
where the fundamental variables satisfy
\begin{equation}
\{z_i, \bar{z}_j\} = i\delta_{ij}, \qquad \{z_i, z_j\} = \{\bar{z}_i, \bar{z}_j\} = 0 \label{eq:zbracket}
\end{equation}
and the matrix $\beta$ is constrained to be symmetric and unitary, $\beta^T = \beta$, $\beta\beta^\dagger = \mathds{1}$, which guarantees compatibility with the Poisson algebra~\eqref{eq:1Poisson}. The variables $z_i$, $\bar{z}_i$ are classical complex phase-space coordinates forming $N$ independent canonical pairs; they arise from the well-known bosonic realization of $\mathfrak{su}(1,1)$~\cite{LL93,LS93,FF97} and can be viewed as the classical counterparts of the annihilation and creation operators $a_i$, $a_i^\dagger$ of a set of harmonic oscillators\footnote{Each pair $(z_i,\bar{z}_i)$ parametrizes a copy of $\mathbb{C}$ with the canonical symplectic structure; the whole phase space is $\mathbb{C}^N$.}.

The isomorphism between the Poisson algebra~\eqref{eq:1Poisson} and the $\mathfrak{su}(1,1)$ Lie algebra induces an isomorphism between the group of bosonic transformations on $(\bar{z},z)$ and the group $SU(1,1)$. This can be seen explicitly by mapping the bosonic variables to canonical ones via $(\bar{z},z) \to (q,p)$ with $q = \frac{1}{\sqrt{2}}(z+\bar{z})$, $p = \frac{1}{\sqrt{2} i}(z-\bar{z})$, so that $\{q,p\}=1$~\cite{LM12}. The evolution of observables in the bosonic realization then takes the form
\begin{align}
V(\varphi) = V(\mathbf{z}(\varphi_0)) \cosh(\tilde{\delta}(\varphi-\varphi_0))\,, \quad
J_+(\varphi) = V(\mathbf{z}(\varphi_0)) \sinh(\tilde{\delta}(\varphi-\varphi_0))\,.
\label{eq:vol-sol}
\end{align}
At $\varphi = \varphi_0$, $J_+$ vanishes and $V$ attains its minimum, exactly as in~\eqref{eq:solns}.

\subsection{Effective FLRW from harmonic cosmology}

Interpreting $V(\varphi)$ in~\eqref{eq:solns} (or~\eqref{eq:vol-sol}) as the spatial volume of the universe, the solution $V(\varphi)$ exhibits a bounce within harmonic cosmology. To make this explicit we derive an effective Friedmann equation, which parallels the modified scale-factor evolution. In the context of LQC~\cite{BojLQC08,BojLQC15} and GFT~\cite{CalGieOri12,GieOriSin13, CalGieOri14, GieOri14, Ori17-1,Gie15,Gielen16}, the parameter $\varphi$ is identified with the value of a free massless scalar field that serves as a relational internal time. This deparametrization~\cite{AS06,Viqar2012,EWE19} uses a monotonic physical degree of freedom as a clock; the scalar field is not a coordinate time and does not tick at the same rate as proper time $\tau$. Instead, the equation of motion $d\phi/d\tau$ follows from the dynamics and is not constant in general.

Choosing $\phi$ as internal time, we consider a free massless scalar field with vanishing potential, so that the scalar Hamiltonian is simply
\begin{equation}
h_\phi  = \frac{1}{2} \frac{p^2_{\phi}}{V},
\label{eq:scalarH}
\end{equation}
and its momentum $p_\phi$ Poisson-commutes with $h_\phi$, hence is a constant of motion. The system can therefore be deparametrized using $\phi$ as internal time and $p_\phi$ as the physical Hamiltonian that generates translations in $\phi$. To match the coefficients that appear in the Friedmann equation, we identify
\begin{equation}
p_\phi \equiv \sqrt{12 \pi G}  H^{\tilde{\delta}}_{\varphi}\,.
\label{eq:identification-pH}
\end{equation}
Inserting the solution~\eqref{eq:solns} for $V(\varphi)$ together with this identification yields the matter density
\begin{equation}
\rho_\phi(\varphi) = \frac{1}{2} \frac{p^2_\phi}{V^2}
= \frac{3}{(8 \pi G) \tilde{\delta}^2} \,
\frac{p^2_\phi}{\big(p^2_\phi+ \frac{3 V^2_{\mathrm{min}}}{(4 \pi G) \tilde{\delta}^2}\big)} \,
\frac{1}{\cosh^2(4 \pi G \tilde{\delta}(\varphi - \varphi_0))}\,,
\end{equation}
where $\tilde{\delta}$ can be rescaled as $\tilde{\delta} \to 4 \pi G \, \delta$. The density attains a finite maximum at the bounce $\varphi = \varphi_0$,
\begin{equation}
(\rho_\phi)_{\mathrm{max}} := \rho_\phi(\varphi_0)
= \bigg(\frac{p^2_\phi}{p^2_\phi+ \frac{3 V^2_{\mathrm{min}}}{(4 \pi G) \delta^2}}\bigg) \, \rho_c \,,
\qquad \rho_c \equiv \frac{3}{(8 \pi G) \delta^2} \,.
\end{equation}
Thus $(\rho_\phi)_{\mathrm{max}}$ depends on both the scalar momentum $p_\phi$ (the dynamical contribution) and the kinematical minimal volume $V_{\mathrm{min}}$. In the limit $p_\phi^2 \gg \frac{3}{4\pi G \delta^2} V^2_{\mathrm{min}}$, the maximum density approaches the critical density $\rho_c$, which is of Planck order when $\delta^2 \propto a_{\mathrm{Pl}} \equiv 8\pi G$.\footnote{We work in units $c=1$, $\hbar=1$.} In LQC, $\delta^2$ is associated with the area spectrum~\cite{AS11}; in GFT~\cite{Oriti16-2,Gielen23} it originates from the microscopic dynamics, setting the fundamental scale (typically the Planck scale). Hence the same algebraic framework underpins the bounce in both approaches, and $\delta^2$ (or $\rho_c$) marks the scale where quantum gravity effects become dominant. Note that for $\mathcal{C}=0$ (the standard scenario in the LQC phase space), the maximum density becomes independent of $p_\phi$ and equals exactly $\rho_c$.

The effective Friedmann equation is obtained by first writing the equation of motion for $\phi$ in proper time $\tau$,
\begin{equation}
\frac{d \phi}{d \tau} = \{\phi, h_\phi \} = \frac{p_\phi}{V}\,.
\end{equation}
Introducing the scale factor $a$ with $V = a^3$, the Hubble rate is
\begin{equation}
\bigg(\frac{1}{a} \frac{d a}{d \tau}\bigg)^2
= \bigg(\frac{1}{3 V} \frac{d \phi}{d \tau}\bigg)^2 \,
\bigg(\frac{d V}{d \phi}\bigg)^2
= \frac{p^2_\phi}{9 V^4} \, \bigg(\frac{d V}{d \phi}\bigg)^2\,,
\label{eq:Friedmann1}
\end{equation}
where the $\phi$-derivative of $V$ follows from the equations of motion~\eqref{eq:eqm-harmonic} together with the identification~\eqref{eq:identification-pH}:
\begin{equation}
\bigg(\frac{d V}{d \phi}\bigg)^2
= 12 \pi G \, \{V, H^{\delta}_{\varphi}\}^2
= 12 \pi G \, J^2_+
= 12 \pi G \, \bigg(V^2 - \frac{4 \pi G}{3} \delta^2 p^2_\phi - V^2_{\mathrm{min}}\bigg)\,.
\label{eq:solV}
\end{equation}
Combining~\eqref{eq:Friedmann1} and~\eqref{eq:solV} we obtain
\begin{equation}
\bigg(\frac{1}{a} \frac{d a}{d \tau}\bigg)^2
= \frac{4 \pi G \, p^2_\phi}{3 V^4}
\bigg(V^2 - \frac{4 \pi G}{3} \delta^2 p^2_\phi - V^2_{\mathrm{min}}\bigg)
= \frac{8 \pi G}{3} \rho_\phi
\bigg(1- \frac{\rho_\phi}{\rho_c} - \frac{V^2_{\mathrm{min}}}{V^2}\bigg)\,.
\label{eq:Friedmann2}
\end{equation}
When the Casimir $\mathcal{C}$ (or equivalently $V_{\mathrm{min}}$) vanishes, Eq.~\eqref{eq:Friedmann2} coincides with the effective dynamics of LQC.

The classical spatially-flat Friedmann equation is recovered in the limit $\tilde{\delta} \to 0$ together with $\mathcal{C}=0$~\cite{BBC19}:
\begin{equation}
\lim_{\tilde{\delta} \to 0} \left(\frac{1}{a} \frac{d a}{d \tau}\right)^2\bigg|_{\mathcal{C}=0}
= \frac{8 \pi G}{3} \rho_\phi .
\label{eq:classicalF}
\end{equation}
In this limit the volume $V$ grows exponentially without a bounce, and the canonical momentum conjugate to $V$ can be identified as $\frac{1}{4 \pi G}\frac{d a}{a d \tau}$. At the algebraic level, $\tilde{\delta}\to 0$ and $\mathcal{C}=0$ imply $J_-=0$ and $\{V,H^0_{\varphi}\}=V$. The condition $\mathcal{C}=0$ (hence $V_{\mathrm{min}}=0$) can also be motivated by the requirement that the volume remains real, because the solution for $V$ would otherwise become imaginary when $\cosh(\sqrt{12\pi G}\,\varphi) > 1$. In this sense $\mathcal{C}=0$ acts as a reality condition. Thus the singularity appearing in the standard Friedmann model is resolved in harmonic cosmology through a bounce.

\section{Multi-mode squeezed states in harmonic cosmology}\label{sec:quantum}

The classical harmonic cosmology reviewed in Sec.~\ref{SII} resembles a set of harmonic oscillators because of its $\mathfrak{su}(1,1)$ algebraic structure. In the multimode bosonic realization of harmonic cosmology (cf.~\eqref{9}), we can quantize the system similarly to the quantization of multiple harmonic oscillators. In other words, we can apply the group-theoretical quantization rule which lifts the bosonic variables $(\bar{z},z)$ to creation and annihilation operators $(a^\dagger,a)$. For the multimode bosonic realization of the $\mathfrak{su}(1,1)$ algebra in \eqref{9}, the same quantization yields the operators
\begin{align}\label{20}
\hat{V} &= \frac{\tilde{\delta}}{4} \sum_{i=1}^N \left( a_i^\dagger a_i + a_i a_i^\dagger \right) = \frac{\tilde{\delta}}{2} \sum_{i=1}^N \left( a_i^\dagger a_i + \frac{1}{2} \right)\,, \\ \nonumber
\hat{J} &= \frac{\tilde{\delta}}{2} \sum_{i,j=1}^N \beta_{ij} a_i^\dagger a_j^\dagger, \qquad 
\hat{J}^\dagger = \frac{\tilde{\delta}}{2} \sum_{i,j=1}^N \bar{\beta}_{ij} a_i a_j \,.
\end{align}
The creation and annihilation operators satisfy the canonical commutation relations $[a_i,a_j^\dagger]=\delta_{ij}$, and it is easy to check that the quantized operators still form an $\mathfrak{su}(1,1)$ algebra:
\begin{align}\label{21}
[\hat{V},\hat{J}]=  \tilde{\delta} \hat{J}\,, \quad [\hat{V}, \hat{J}^{\dagger}] = -\tilde{\delta}\hat{J}^{\dagger}\,, \quad [\hat{J},\hat{J}^{\dagger}]= -2 \tilde{\delta} \hat{V}.
\end{align}

The multimode bosonic realization \eqref{9} and the resulting quantum operators \eqref{20} also appear in group field theory (GFT) condensate cosmology~\cite{CalGieOri14, GieOri14, Ori17-1,Gie15,Gielen16}, in terms of ladder operators $a_i^\dagger, a_i$, where the mode index $i$ labels a single-particle basis. In that setting, the fundamental GFT field is expanded in a discrete set of modes, and the quantized theory is built upon $a_i$ and $a_i^\dagger$, which create and annihilate quanta of geometry---the microscopic building blocks of the emergent spacetime. The volume operator $\hat{V}$ is proportional to the total number of quanta, given by the sum over all modes of the individual mode number operators $\sum_i a_i^\dagger a_i$ (plus a constant zero-point term). Thus, its expectation value $\langle\hat{V}\rangle$ directly measures the total number of quanta in the condensate, not to be confused with the number $N$ of independent modes, which is a fixed parameter of the model. The quadratic operators $\hat{J}$ and $\hat{J}^\dagger$ create and annihilate correlated pairs of quanta and are precisely the building blocks of the squeezing operator; indeed, the Hamiltonian $H^{\tilde{\delta}}_{\varphi} \propto \hat{J}_-$ of harmonic cosmology~\eqref{eq:hamiltonian} is a linear combination of these generators and acts as a squeezing Hamiltonian.

Given the $\mathfrak{su}(1,1)$ algebra \eqref{21}, we can define the algebraic displacement operator as $D(\alpha)=\exp\{\alpha \hat{J}-\alpha^*\hat{\bar{J}}\}$ with $\alpha\in\mathbb{C}$. When acting on the vacuum state $\ket{0,k}$ defined by $\hat{\bar{J}}\ket{0,k}=0$ and $\hat{V}\ket{0,k}=k\ket{0,k}$ with Bargmann index $k$ (in the positive discrete series of the unitary irreducible representation of $\mathfrak{su}(1,1)$), we obtain the Perelomov generalized $SU(1,1)$ coherent state $D(\alpha)\ket{0,k}$. As observed in \cite{LL93,LS93}, owing to the multimode bosonic realization \eqref{20}, the generalized $\mathfrak{su}(1,1)$ coherent states coincide with the multimode squeezed states:
\begin{equation}
S(\beta)\ket{0}=\exp\Bigl\{\frac{\tilde{\delta}}{2}\sum_{i,j}(\beta_{ij}a_i^\dag a_j^\dag- \bar{\beta}_{ij}a_j a_i)\Bigr\}\ket{0}
\label{eq:squeezed-operator}
\end{equation}
where $\ket{0}$ is the Fock vacuum, $a_i\ket{0}=0$ for all $i$, which is consistent with $\hat{J}^{\dagger}\ket{0,k}=0$ when the representation is appropriately matched, and $\beta_{ij}$ are parameters encoding the squeezing. Here, the multimode squeezing unitary operator $S(\beta)$ ($S S^{\dagger}= I$) generates a squeezing transformation, i.e.\ a Bogoliubov transformation, of the ladder operators:
\begin{equation}
S^\dag a_i S = \sum_{j=1}^N (\mu_{ij} a_j + \nu_{ij} a_j^\dagger),\qquad
S^\dag a_i^\dagger S = \sum_{j=1}^N (\bar{\mu}_{ij} a_j^\dagger + \bar{\nu}_{ij} a_j),
\label{eq:bogo}
\end{equation}
where the matrices $\mu$ and $\nu$ satisfy $\mu\mu^\dagger - \nu\nu^\dagger = \mathds{1}$ and $\mu\nu^T = \nu\mu^T$. Their entries are related to the matrix $(\tilde{\delta}\beta_{ij})\equiv{\bm{\beta}}$ by~\cite{LL93}
\begin{equation}
\mu_{ij}=(\cosh\bm{\beta})_{ij},\quad \nu_{ij}=(\sinh\bm{\beta})_{ij}.
\label{eq:relation-beta-nu-mu}
\end{equation}

The multimode squeezed states constructed here are the natural generalization of the dipole condensate states studied in GFT cosmology, where single-mode squeezing has been shown to produce large quantum fluctuations of the volume. By extending the analysis to many modes, we place these states within the broader framework of $\mathfrak{su}(1,1)$ harmonic cosmology and provide the physical interpretation of the squeezing parameters as encoding correlations between pairs of GFT quanta~\cite{Calcinari:2026hbm}. This connection not only motivates the algebraic structure of the present model but also ensures that our results are directly applicable to the semiclassical analysis of GFT condensates.

We stress that the transformation \eqref{eq:bogo} and the squeezed state $S(\beta)\ket{0}$ constructed here are kinematical: they define a class of states in the full Fock space without imposing any dynamical constraint. The Hamiltonian $H^{\tilde{\delta}}$ also generates a squeezing-type transformation when it acts as the time-evolution operator, but this dynamical process is not the subject of the present analysis. Here we focus exclusively on the kinematical properties of multimode squeezed states, specifically, the expectation values and uncertainties of the volume operator explicitly computed in the next section.

\subsection{Semiclassicality of multimode squeezed states} 
\label{sec:semiclassical}

Conventionally, semiclassicality is assessed through the relative uncertainties of the relevant observables in the physical system of interest. Thus, a small relative uncertainty indicates that the state is sharply peaked and can be approximated by a classical trajectory. Here, we compute the relative uncertainty of the volume operator $\hat{V}$,
\begin{equation}
\frac{\Delta \hat{V}}{\langle \hat{V} \rangle} = \frac{\sqrt{\langle \hat{V}^2 \rangle - \langle \hat{V} \rangle^2}}{\langle \hat{V} \rangle}. \label{eq:relunc}
\end{equation}

For the multimode squeezed vacuum $|\Psi \rangle = S(\mu,\nu)|0\rangle$, we need to compute the expectation values of $\hat{V}$ and $\hat{V}^2$ (we denote the expectation value $\langle \Psi | \hat{A} | \Psi \rangle$ by $\langle \hat{A} \rangle_{\Psi}$). Using the Bogoliubov transformation \eqref{eq:bogo}, we find
\begin{equation}
\langle a_i^\dagger a_j \rangle_{\Psi} = \sum_k \bar{\nu}_{ik} \nu_{jk} = (\nu^\dagger \nu)_{ij}, \qquad
\langle a_i a_j \rangle_{\Psi} = \sum_k \mu_{ik} \nu_{jk} = (\mu \nu^T)_{ij}. \label{eq:expect}
\end{equation}
The expectation value of the volume operator is then
\begin{equation}
\langle \hat{V} \rangle_{\Psi} = \frac{\tilde{\delta}}{2} \sum^N_{i=1} \left( \langle a_i^\dagger a_i \rangle_{\Psi} + \frac{1}{2} \right) = \frac{\tilde{\delta}}{2} \left( \operatorname{tr}(\nu^\dagger \nu) + \frac{N}{2} \right). \label{eq:meanV}
\end{equation}

To compute $\langle \hat{V}^2 \rangle_{\Psi}$, we note that $\hat{V}^2 = \frac{\tilde{\delta}^2}{4} \sum_{i,j} \left( a_i^\dagger a_i + \frac{1}{2} \right) \left( a_j^\dagger a_j + \frac{1}{2} \right)$. Using Wick's theorem for Gaussian states~\cite{SP02}, we obtain
\begin{align}
\langle a_i^\dagger a_i a_j^\dagger a_j \rangle_{\Psi} &= \langle a_i^\dagger a_i \rangle_{\Psi} \langle a_j^\dagger a_j \rangle_{\Psi} + \langle a_i^\dagger a_j \rangle_{\Psi} \langle a_i a_j^\dagger \rangle_{\Psi} + \langle a_i^\dagger a_j^\dagger \rangle_{\Psi} \langle a_i a_j \rangle_{\Psi} \nonumber \\
&= \langle n_i \rangle_{\Psi} \langle n_j \rangle_{\Psi} + |\langle a_i^\dagger a_j \rangle_{\Psi}|^2 + |\langle a_i a_j \rangle|_{\Psi}^2, \label{eq:wick}
\end{align}
where $n_i \equiv a_i^\dagger a_i$ is the number operator and we used $a_i a_j^\dagger = a_j^\dagger a_i + \delta_{ij}$ with $\langle a_i a_j^\dagger \rangle_{\Psi} = \delta_{ij} + \langle a_j^\dagger a_i \rangle_{\Psi}$.

Summing over $i,j$ and including the constant terms, $\langle \hat{V}^2 \rangle_{\Psi}$ is given by
\begin{equation}
\langle \hat{V}^2 \rangle_{\Psi} = \frac{\tilde{\delta}^2}{4} \left[ \left( \sum_{i=1}^N \langle n_i \rangle_{\Psi} + \frac{N}{2} \right)^2 + \sum^N_{i=1} \langle n_i \rangle_{\Psi} + \sum^N_{i,j} \left( |\langle a_i^\dagger a_j \rangle_{\Psi}|^2 + |\langle a_i a_j \rangle|_{\Psi}^2 \right) \right]. \label{eq:V2}
\end{equation}
The variance is then
\begin{equation}
(\Delta \hat{V}^2)_{\Psi} = \langle \hat{V}^2 \rangle_{\Psi} - \langle \hat{V} \rangle_{\Psi}^2 = \frac{\tilde{\delta}^2}{4} \left[ \sum^N_{i=1} \langle n_i \rangle_{\Psi} + \sum^N_{i,j} \left( |\langle a_i^\dagger a_j \rangle_{\Psi}|^2 + |\langle a_i a_j \rangle_{\Psi}|^2 \right) \right]. \label{eq:varV}
\end{equation}

This expression simplifies considerably in the case where the squeezing matrix $\beta$ is diagonal, $\beta_{ij} = \beta_i \delta_{ij}$ with $\beta_i$ real. Imposing such a diagonal squeezing matrix corresponds to squeezing that acts independently on each mode, producing pairs of quanta that belong to the same mode, so that all inter-mode correlations ($i\neq j$) are neglected. This simplification is not completely arbitrary: the volume operator \eqref{20} is diagonal in the chosen mode basis, and the diagonal squeezing configuration is precisely the one that aligns the squeezing with the eigenbasis of $\hat{V}$, and is therefore expected to minimize the relative volume uncertainty. In this sense, the diagonal case can be regarded as the most natural semiclassical configuration for a given average volume.

Now the squeezing operator factorizes into a product of single-mode squeeze operators, and each mode is squeezed independently. The Bogoliubov coefficients for mode $i$ are $\mu_i = \cosh \beta_i$, $\nu_i = \sinh \beta_i$, using the relation in Eq.~\eqref{eq:relation-beta-nu-mu} and absorbing $\tilde{\delta}$ into the definition of $\beta_i$. Then $\langle n_i \rangle_{\Psi} = \sinh^2 \beta_i$, $\langle a_i^\dagger a_j \rangle_{\Psi} = 0$ for $i \neq j$, and $\langle a_i a_j \rangle_{\Psi}= \delta_{ij} \sinh \beta_i \cosh \beta_i$. Consequently,
\begin{equation}
\langle \hat{V} \rangle_{\Psi} = \frac{\tilde{\delta}}{2} \sum^N_{i=1} \left( \sinh^2 \beta_i + \frac{1}{2} \right), \qquad
(\Delta \hat{V}^2)_{\Psi} = \frac{\tilde{\delta}^2}{4} \sum^N_{i=1} \left( \sinh^2 \beta_i + 2\sinh^2 \beta_i \cosh^2 \beta_i \right). \label{eq:diag}
\end{equation}
The relative uncertainty of the volume becomes
\begin{equation}
\frac{(\Delta \hat{V})_{\Psi}}{\langle \hat{V} \rangle_{\Psi}} = \frac{ \sqrt{ \sum \limits^N_{i=1} \left( \sinh^2 \beta_i + 2\sinh^2 \beta_i \cosh^2 \beta_i \right) } }{ \sum \limits^N_{i=1} \left( \sinh^2 \beta_i + \frac{1}{2} \right) }. \label{eq:rel_diag}
\end{equation}
If all modes have the same squeezing $\beta_i = \beta$, then 
\begin{equation}
\frac{(\Delta \hat{V})_{\Psi}}{\langle \hat{V} \rangle_{\Psi}} = \frac{ \sqrt{N \left( \sinh^2 \beta + 2\sinh^2 \beta \cosh^2 \beta \right) } }{ N \left( \sinh^2 \beta + \frac{1}{2} \right) } = \frac{ \sqrt{\sinh^2 \beta + 2\sinh^2 \beta \cosh^2 \beta} }{ \sqrt{N} \left( \sinh^2 \beta + \frac{1}{2} \right) } \,. 
\label{eq:rel_equal}
\end{equation}

It should be emphasized that $N$ denotes the number of independent modes in the bosonic realization, not the number of quanta. Each mode can be populated by many quanta, and the total number of quanta is $\sum_i \langle n_i \rangle$. For a single mode ($N=1$), Eq.~\eqref{eq:rel_equal} reduces to the large-uncertainty result known from the single-mode squeezed vacuum state in GFT~\cite{GP20,CG23}. Our multimode analysis thus directly generalizes those works: the relative uncertainty is suppressed by a factor $1/\sqrt{N}$ when all modes are squeezed equally.

\begin{figure}[htbp]
    \centering
    \subfloat[$\Delta \hat{V} / \langle \hat{V} \rangle$ as a function of $\beta$ for fixed sets of $N$.]{%
        \includegraphics[width=0.48\textwidth]{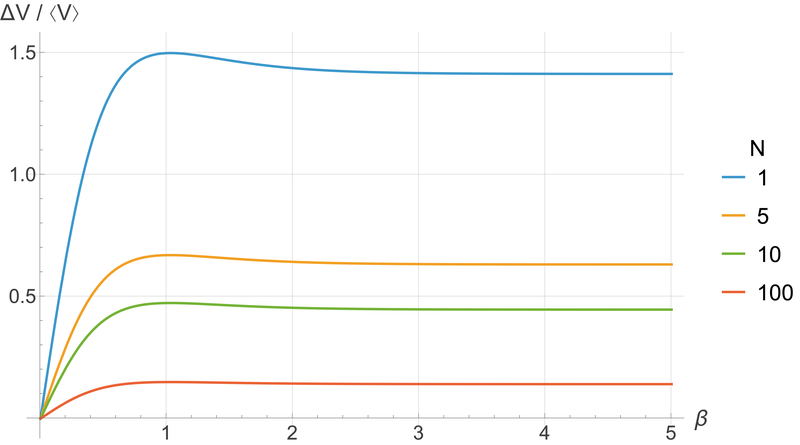}\label{fig:uncertainty_N1}}
    \hfill
    \subfloat[$\Delta \hat{V} / \langle \hat{V} \rangle$ as a function of $N$ for fixed sets of $\beta$.]{%
        \includegraphics[width=0.48\textwidth]{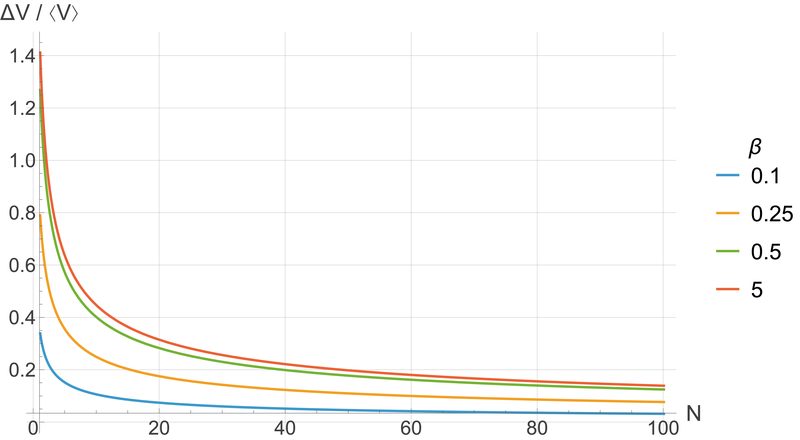}\label{fig:uncertainty_Nlarge}}
    \caption{Relative uncertainty $\Delta \hat{V} / \langle \hat{V} \rangle$.}
    \label{fig:uncertainty_combined}
\end{figure}

Fig.~\ref{fig:uncertainty_combined} shows the behavior of the relative uncertainty of the volume operator in the squeezed vacuum as functions of the squeezing parameter and the number of modes. For single-mode ($N=1$) squeezing ($\beta > 0$), a large relative uncertainty appears and saturates for $\beta > 1$. For large squeezing $\beta \gg 1$, $\sinh \beta \sim \cosh \beta \sim \frac{1}{2} e^\beta$, so $\sinh^2 \beta \sim \frac{1}{4} e^{2 \beta}$ and $\sinh^2 \beta \cosh^2 \beta \sim \frac{1}{16} e^{4 \beta}$. Then the numerator in Eq.~\eqref{eq:rel_equal} scales as $\sqrt{ \frac{1}{16} e^{4 \beta} } = \frac{1}{4} e^{2 \beta}$, while the denominator scales as $ \sqrt{N} \frac{1}{4} e^{2 \beta}$, giving $\frac{1}{\sqrt{N}}$. Thus for fixed $\beta$, the relative uncertainty decreases as $1/\sqrt{N}$ with the number of modes $N$ (see Fig.~\ref{fig:uncertainty_combined}). This yields the following key result: a multimode squeezed vacuum can be made semiclassical by taking a large number of modes, even without displacement.

If the modes are not equally squeezed, the sum over modes still yields a factor of $N$ in the denominator, provided the squeezing parameters are of similar magnitude. Hence, for a condensate with many modes, the state becomes sharply peaked in volume.

We note that the above calculation assumed a diagonal squeezing matrix. In the general case, off-diagonal squeezing (or mode mixing) introduces additional correlations that may affect the variance. However, since $\hat{V}$ is invariant under unitary mode mixing, the expectation $\langle \hat{V} \rangle_{\Psi}$ remains unchanged after diagonalization (see Eq.~\eqref{eq:meanV}). The variance may receive extra contributions from off-diagonal terms, but these are positive and can only increase the relative uncertainty. Therefore, the diagonal case provides a lower bound on semiclassicality; any mode mixing makes the state less semiclassical for fixed squeezing magnitudes. This suggests that semiclassicality can be optimally achieved when the squeezing is aligned with the mode basis that diagonalizes $\hat{V}$. Connecting to GFT, in a single-mode condensate, the relative uncertainty remains of order unity, and additional mechanisms (displacement, mixing, or thermal effects) are required to obtain a semiclassical geometry~\cite{GP20,CG23}. The diagonal multimode case studied here isolates the minimal fluctuation scenario and provides a transparent understanding of how the $1/\sqrt{N}$ suppression of independent fluctuations can render a purely squeezed vacuum semiclassical when the number of independent modes is large.

\section{Summary and discussion}
\label{sec:conclusion}

In this paper, we have studied multimode squeezed states within the framework of harmonic cosmology, which is based on the $\mathfrak{su}(1,1)$ Lie algebra and provides a unified description of bouncing cosmologies in loop quantum cosmology and group field theory (GFT). We constructed the multimode bosonic realization of $\mathfrak{su}(1,1)$ given by the operators $\hat{V}$, $\hat{J}$, and $\hat{J}^\dagger$ defined in Eq.~\eqref{21}. The multimode squeezing operator was introduced in Eq.~\eqref{eq:squeezed-operator}; by means of the Bogoliubov transformation $b_k = \sum_j (u_{kj} a_j + v_{kj} a_j^\dagger)$ we computed the expectation value $\langle \hat{V} \rangle$ and the uncertainty $\Delta \hat{V}$ in the multimode squeezed vacuum state $\ket{\beta}= S(\beta) \ket{0}$. The relative uncertainty of the volume operator~\eqref{eq:relunc} was evaluated analytically in the diagonal squeezing basis, yielding expressions that depend explicitly on the squeezing parameters and on the number of modes $N$.

Our analysis has been carried out at the kinematical level of the quantum theory; that is, we have employed the full Fock space and the operator algebra without imposing the Hamiltonian constraint that generates the dynamics of harmonic cosmology. The squeezed states considered here therefore serve as a well-defined arena for investigating the emergence of semiclassical spacetime geometry at the level of kinematical observables. This restriction to kinematical states is common in the initial study of semiclassicality in quantum cosmology and allows one to isolate the statistical properties of geometry operators before dealing with the subtleties of the theory's dynamics.

Our main finding is that the relative uncertainty $\Delta \hat{V} / \langle \hat{V} \rangle$ can become small when the number of modes $N$ is large, even in the absence of displacement. Specifically, for a uniform squeezing configuration, the relative uncertainty scales as $1/\sqrt{N}$ at large $N$, in marked contrast to the single-mode case ($N=1$), where it approaches unity for strong squeezing. This behavior arises because the volume operator sums over many modes, and the independent fluctuations of each mode add incoherently, leading to a suppression of the total relative uncertainty via the central limit theorem. Thus, a multimode squeezed vacuum state, representing a condensate of geometry with many independent modes, can behave semiclassically. This result is relevant for understanding how a classical universe emerges from a quantum gravity regime, providing a bridge between the microscopic quantum geometry and the smooth FLRW background. In GFT cosmology, a cosmological universe is often described as a condensate of many quanta that may occupy a single mode; our result shows that a large number of independent modes can also lead to a suppression of quantum fluctuations, an effect that may be realised in extended GFT models.

We also noted that mode mixing (off-diagonal squeezing) tends to increase fluctuations, so the most semiclassical states are those where the squeezing is aligned with the mode basis that diagonalizes the volume operator. In the diagonal basis, the relative uncertainty is minimized, highlighting the importance of choosing an appropriate basis for the construction of semiclassical states in quantum cosmology. This observation is tied to the structure of the squeezing matrix and its diagonalization via factorisation, which isolates the independent two-mode squeezing processes.

The present work opens several directions for further investigation. First, one should consider the explicit dynamical evolution of these states under the harmonic cosmology Hamiltonian $H \propto \hat{J}_-$, which generates a bounce in the volume, and relate their evolution to the effective Friedmann equation that emerges in loop quantum cosmology. In particular, one could investigate whether the kinematical semiclassicality observed here is preserved under the dynamical constraint, and how it connects to the full constraint algebra of quantum gravity---specifically the Complexifier-Volume-Hamiltonian (CVH) algebra of LQC~\cite{BL17}. Second, one can enlarge the class of states beyond the squeezed vacuum by including displacement operators, leading to $\mathfrak{su}(1,1)$ Perelomov coherent states or thermal (mixed) squeezed states; such states, as recently studied in GFT cosmology~\cite{AK20, CG23}, could provide a richer description of semiclassical geometry. Third, the multimode squeezed vacuum is an entangled state across the different modes, and it would be illuminating to quantify the role of multipartite entanglement in the suppression of volume fluctuations. Finally, the techniques developed here can be extended to anisotropic or inhomogeneous models, where multiple modes naturally appear, and may provide a bridge between the effective Friedmann dynamics of loop quantum cosmology and the microscopic condensate picture of GFT.

\section*{Acknowledgements} 
BB would like to thank Xiao-Kan Guo for insightful discussions and acknowledges the support from the Scientific and Technological Research Council of T\"urkiye (T\"UB\.{I}TAK) 3501 under project no. 125F279. 

\appendix
\section{Multimode bosonic realizations of $\mathfrak{su}(1,1)$ algebra}\label{app:multimode}

We provide details on the multimode bosonic realization (\ref{9}). Let $z_i$, $\bar{z}_i$ satisfy $\{z_i, \bar{z}_j\} = i\delta_{ij}$. Define
\[
V = \frac{\delta}{4} \sum_i \bar{z}_i z_i + z_i \bar{z}_i,\quad J = \frac{\delta}{2} \sum_{i,j} \beta_{ij} \bar{z}_i \bar{z}_j,\quad \bar{J} = \frac{\delta}{2} \sum_{i,j} \bar{\beta}_{ij} z_i z_j.
\]
We require that these satisfy the Poisson algebra (\ref{eq:1Poisson}). Compute
\[
\{V, J\} = \frac{\delta^2}{8} \sum_{i,j,k} \beta_{jk} \{ \bar{z}_i z_i + z_i \bar{z}_i, \bar{z}_j \bar{z}_k \}.
\]
Using $\{z_i,\bar{z}_j\}=i\delta_{ij}$ and $\{\bar{z}_i, \bar{z}_j\} = 0$, we obtain 
\[
\{ \bar{z}_i z_i, \bar{z}_j \bar{z}_k \} = \bar{z}_i \{z_i, \bar{z}_j \bar{z}_k\} + \{ \bar{z}_i, \bar{z}_j \bar{z}_k\} z_i = \bar{z}_i ( \{ z_i, \bar{z}_j\} \bar{z}_k + \bar{z}_j \{z_i, \bar{z}_k\} ) = i\delta_{ij} \bar{z}_i \bar{z}_k + i\delta_{ik} \bar{z}_i \bar{z}_j \,,
\]
and similarly, 
\[
\{ z_i \bar{z}_i, \bar{z}_j \bar{z}_k \} = z_i \{\bar{z}_i, \bar{z}_j \bar{z}_k\} + \{ z_i, \bar{z}_j \bar{z}_k\} \bar{z}_i = ( \{ z_i, \bar{z}_j\} \bar{z}_k + \bar{z}_j \{z_i, \bar{z}_k\} ) \bar{z}_i = i\delta_{ij} \bar{z}_i \bar{z}_k + i\delta_{ik} \bar{z}_i \bar{z}_j \,.
\]
Thus
\[
\{V, J\} = \frac{\delta^2}{8} \sum_{i,j,k} \beta_{jk} (2 i) ( \delta_{ij} \bar{z}_k \bar{z}_i + \delta_{ik} \bar{z}_j \bar{z}_i ) = i \frac{\delta^2}{4} \left( \sum_{i,k} \beta_{ik} \bar{z}_k \bar{z}_i + \sum_{i,j} \beta_{ji} \bar{z}_j \bar{z}_i \right).
\]
The matrix $\beta$ is constrained to be symmetric, $\beta_{ik} = \beta_{ki}$, so that the two sums are equal, and giving
\[
\{V, J\} = i \frac{\delta^2}{2} \sum_{i,k} \beta_{ij} \bar{z}_i \bar{z}_j = i \delta J.
\]
where the bracket $\{V,J\} = i\delta J$ satisfied according to (\ref{eq:1Poisson}). Similarly, one finds $\{J,\bar{J}\} = -2i\delta V$ provided that $\beta\beta^\dagger = \mathds{1}$. The condition $\beta\beta^\dagger = \mathds{1}$ ensures that the map $z \to \beta z$ is unitary, preserving the Poisson structure.

\end{document}